\documentclass[cernpreprint,texlive=2020,UKenglish,texmf]{atlasdoc}

\usepackage[subcaption]{atlaspackage}
\usepackage{tabularx}
\usepackage{multirow}
\usepackage{makecell}
\usepackage{atlasbiblatex}
 
\usepackage{atlasphysics}
 
\graphicspath{{logos/}{figures/}}
 
\usepackage{ANA-EXOT-2018-36-PAPER-defs}

\AtlasTitle{Search for charged-lepton-flavour violation in $Z$-boson decays with the ATLAS detector}

\AtlasAbstract{
Leptons with essentially the same properties apart from their mass are grouped into three families (or flavours). The number of leptons of each flavour is conserved in interactions, but this is not imposed by fundamental principles. Since the formulation of the standard model of particle physics, the observation of flavour oscillations among neutrinos has shown that lepton flavour is not conserved in neutrino weak interactions. So far, there has been no experimental evidence that this also occurs in interactions between charged leptons. Such an observation would be a sign of undiscovered particles or a yet unknown type of interaction. Here the ATLAS experiment at the Large Hadron Collider at CERN reports a constraint on lepton-flavour-violating effects in weak interactions, searching for $Z$-boson decays into a $\tau$ lepton and another lepton of different flavour with opposite electric charge. The branching fractions for these decays are measured to be less than $8.1\times10^{-6}$ ($e\tau$) and $9.5\times10^{-6}$ ($\mu\tau$) at 95\% confidence level
using 139 fb$^{-1}$ of proton--proton collision data at a centre-of-mass energy
of $\sqrt{s}=13$ TeV and 20.3 fb$^{-1}$ at $\sqrt{s}=8$ TeV.  These results
supersede the limits from the Large Electron–Positron Collider experiments
conducted more than two decades ago.
}

\AtlasRefCode{EXOT-2018-36}
 
\PreprintIdNumber{CERN-EP-2020-163}

\AtlasJournal{Nature Physics}
\AtlasJournalRef{Nature Physics 17 (2021) 819–825}
\AtlasDOI{10.1038/s41567-021-01225-z}
\hypersetup{pdftitle={ATLAS document},pdfauthor={The ATLAS Collaboration}}

\begin{document}
 
\maketitle

\clearpage

In the standard Model of particle physics (SM) \cite{SM1,SM2,SM3,SM4}, three lepton families (flavours) exist.
The number of leptons of each family is conserved in weak interactions, and violation of this assumption is known as lepton flavour violation (LFV).
No fundamental principles forbid LFV processes in the SM.
The phenomenon of neutrino oscillations, where neutrinos (the neutral leptons) of one flavour transform into those of another \cite{NO1,NO2},
indicates that neutrinos have mass and LFV processes do occur in nature.
The mechanisms responsible for neutrinos acquiring mass and weak interactions violating lepton flavour conservation remain unknown.
More experimental data are needed to constrain and guide possible generalizations of the SM
explaining these phenomena.
 
An observation of LFV in charged-lepton interactions would be an unambiguous sign of new physics.
In particular, decays of the $Z$ boson into a light lepton (electron or muon) and a $\tau$ lepton at colliders are of experimental interest.
The abundance of $Z$ bosons produced at the Large Hadron Collider (LHC) offers the opportunity to strongly constrain potential LFV \Zet or \Zmt interactions,  in particular those proportional to the centre-of-mass energy of the decay~\cite{Davidson2012}.  Moreover, the $Z\to e\tau,\mu\tau$ decays are less constrained by low-energy experiments than \Zem decays.
According to current knowledge, these decays can occur via neutrino mixing but are too rare to be detected.
Only 1 in approximately $10^{54}$ $Z$ bosons would decay into a muon and a $\tau$ lepton \cite{ZLFV-predictions}.
An observation of such decays would therefore require new theoretical explanations.
For example, theories predicting the existence of heavy neutrinos \cite{HN} provide a fundamental understanding of the observed tiny masses and large mixing of the SM neutrinos. In such theories, up to 1 in $10^{5}$ $Z$ bosons would be expected to undergo an LFV decay involving $\tau$ leptons.
The ATLAS experiment can test the predictions of such theories by observing or setting ever more stringent constraints on LFV $Z$-boson decays.
 
Constraints on the branching fractions ($\BR$) of the LFV decays of the $Z$ boson involving a $\tau$ lepton have been set by the experiments at the Large Electron-Positron Collider (LEP): $\BR(\Zet) < 9.8\times10^{-6}$~\cite{LEP-Opal}
and $\BR(\Zmt) < 1.2\times10^{-5}$~\cite{LEP-Delphi}  at 95\% confidence level (CL).
The ATLAS experiment~\cite{PERF-2007-01} at the LHC has set constraints $\BR(\Zet) < 5.8\times10^{-5}$ at 95\% CL using part of the Run~2 data
and $\BR(\Zmt) < 1.3\times10^{-5}$ using the Run~1 data and a subset of the Run 2 data \cite{EXOT-2016-36}.

This work uses proton--proton ($pp$) collision data collected by the ATLAS experiment during Run~2 of the LHC, containing about eight billion $Z$-boson decays.
Only events with a $\tau$ lepton that decays hadronically are considered.
Neural network (NN) classifiers are used in a novel way for optimal discrimination of signal from backgrounds,
and to achieve improved sensitivity in the search for LFV effects in the data using a binned maximum-likelihood fit.
The result for the $\mu\tau$ channel is combined with a previous LHC Run 1 result to further improve the sensitivity.
These results set constraints on LFV $Z$-boson decays involving $\tau$ leptons that supersede the most stringent ones set by the LEP experiments more than two decades ago.

\section{The ATLAS experiment and data sample}
 
To record and analyse the LHC $pp$ collisions, the ATLAS experiment uses a multipurpose particle
detector with a forward--backward symmetric cylindrical geometry and a near
$4\pi$ coverage in solid angle~\cite{PERF-2007-01,ATLAS-TDR-2010-19,PIX-2018-001}.
It consists of an inner tracking detector surrounded by a superconducting
solenoid, electromagnetic and hadronic calorimeters, and a muon spectrometer.
 
The search uses the complete dataset of $pp$ collision events at a centre-of-mass energy $\sqrt{s}= \SI{13}{\TeV}$
collected by the ATLAS experiment during the LHC Run~2.
This dataset was recorded using single-electron or single-muon triggers~\cite{TRIG-2016-01}
and corresponds to an integrated luminosity of \SI{139}{\ifb}.
For the search in the $\mu\tau$ channel, the results are combined with those of a previous similar search using
$pp$ collisions at $\sqrt{s}= \SI{8}{\TeV}$ during the LHC Run~1,
corresponding to an integrated luminosity of \SI{20.3}{\ifb}~\cite{HIGG-2015-09}.
 
Candidates for
electrons \cite{EGAM-2018-01},
muons \cite{PERF-2015-10},
jets \cite{Cacciari:2008gp,Fastjet,JETM-2018-05},
and visible decay products of hadronic $\tau$-lepton decays (\tauhadvis) \cite{ATL-PHYS-PUB-2019-033,ATLAS-CONF-2017-029}
are reconstructed from energy deposits in the calorimeters and charged-particle tracks measured in the inner detector and the muon spectrometer.
 
Electron candidates are required to pass
the \Medium likelihood-based identification requirement \cite{EGAM-2018-01}
and have pseudorapidity 
$\abs{\eta} < 1.37$ or $1.52<\abs{\eta} < 2.47$.
Muon candidates are required to pass the \Medium identification requirement \cite{PERF-2015-10} and have $\abs{\eta} < 2.5$.
Both the electron and muon candidates must have transverse momentum $\pT > \SI{30}{\GeV}$ and satisfy the \Tight isolation requirement \cite{EGAM-2018-01,PERF-2015-10}.
The lower bounds on the electron and muon transverse momenta are driven by the acceptance of the trigger selection.
 
Quark- or gluon-initiated particle showers (jets) are reconstructed using the anti-$k_t$ algorithm \cite{Cacciari:2008gp,Fastjet} with the radius parameter $R = 0.4$.
Jets fulfilling \pT $>20$ {\GeV} and $\abs{\eta} < 2.5$ are identified as
containing $b$ hadrons if tagged by a dedicated multivariate algorithm \cite{FTAG-2018-01}.
 
The \tauhadvis candidates are reconstructed
from jets with $\pT > \SI{10}{\GeV}$, $\abs{\eta} < 1.37$ or $1.52<\abs{\eta} < 2.5$, and one or three associated tracks,
referred to as `1-prong' (1P) and `3-prong' (3P), respectively.
The \tauhadvis identification is performed by a recurrent NN algorithm \cite{ATL-PHYS-PUB-2019-033},
which uses calorimetric shower shapes and tracking information
to discriminate true \tauhadvis candidates from fake candidates from quark- or gluon-initiated jets.
The \tauhadvis candidates are required to pass the \Tight identification selection,
which has an efficiency of 60\% (45\%) for true 1P (3P) \tauhadvis candidates,
constant in the \tauhadvis candidates' transverse momentum,
and a misidentification rate of 1 in 70 (700) for fake 1P (3P) candidates in dijet events.
Dedicated multivariate algorithms are used to further discriminate between \tauhadvis and electrons, and to calibrate the \tauhadvis energy \cite{ATLAS-CONF-2017-029}.
The \tauhadvis candidate with the largest \pT in each event is the selected candidate and is required to have $\pT > \SI{25}{\GeV}$.
Based on simulation, in $Z\to\ell\tau$ decays, the \tauhadvis candidate is expected to be correctly selected 98\% of the time.
 
The missing transverse momentum (\MET) is calculated as the negative vectorial sum of the \pT of all fully reconstructed and calibrated physics objects \cite{PERF-2016-07,ATLAS-CONF-2018-023}.
The calculation also includes inner detector tracks that originate from the vertex associated with the hard-scattering process but are not associated with any of the reconstructed objects.
The missing transverse momentum is the best proxy for the total transverse momentum of undetected particles (in particular neutrinos) in an event.

\section{Search strategy}
 
The $Z\to\ell\tau\to\ell\tauhadvis+\nu$ ($\ell=$ light lepton, $e$ or $\mu$) signal events have a number of
key features that can be exploited to
separate them from the SM background events.
The signal events are
characterized by their unique final state,
which has exactly one $\ell$ and one $\tau$ lepton, with the
invariant mass of the pair being compatible with the $Z$-boson mass.
The $\ell$ and $\tau$ leptons carry opposite electric charges and are emitted
approximately back to back in the plane transverse to the proton beam direction.
Since the $\tau$ lepton is typically boosted due to
the large difference between its mass and the mass of its parent $Z$ boson,
the neutrino from its decay is usually almost collinear with the visible $\tau$-decay products.
The neutrino escapes detection and is reconstructed as part of the \MET of the event.
In a signal event, this is the only major source of \MET.
 
The major background contributions for this search are as follows:
lepton-flavour-conserving $Z\to\tau\tau\to\ell\tauhadvis+3\nu$ decays, where one
of the $\tau$ leptons decays leptonically and the other hadronically;
$Z\to\ell\ell$ decays, where one of the light leptons is misidentified as the
\tauhadvis candidate;
events with a quark- or gluon-initiated jet that is misidentified as the \tauhadvis candidate.
The last of these are hereafter referred to as events with `fakes' and are mostly $W(\to\ell\nu)\text{+jets}$ events and purely hadronic multijet events.
Other SM processes with a real $\ell\tauhadvis$ final state, such
as decays of a top-antitop-quark pair, two gauge bosons or a Higgs boson,
and those with a real \tauhadvis and a jet misidentified as a light lepton, such as $W(\to\tau\nu)\text{+jets}$,
are considered, although their contribution to the overall background is minor.
 
The signal and background events are separated by using a set of event selection
criteria that help to define a signal-enhanced sample, referred to as signal region (SR).
The main selection criteria are listed in Table~\ref{tab:region_defs}
and will be explained in the following.
They are primarily based on the multiplicity of
reconstructed particle candidates and the event topology,
in particular the transverse masses ($\mT$), which are defined as
\begin{equation}\nonumber
\mT(X,\MET) \equiv \sqrt{2 \pT(X) \MET \brkt{1-\cos(\phi_{X}-\phi_{\MET})}},
\end{equation}
where $X$ is either a light lepton or a \tauhadvis candidate and $\phi$ denotes the azimuthal angle.
A schematic of the expected signal and background topologies is described in Extended Data Figures~\ref{fig:mT_dists} and~\ref{fig:mT_dists_2D}.
 
\begin{table}[t]
\caption{Main selection criteria for events in the signal region}
\label{tab:region_defs}
\centering
\begin{tabular}{p{0.48\textwidth}|p{0.48\textwidth}}
\toprule
\textbf{Main selection criteria} & \textbf{Purpose} \\
\toprule
At least one \tauhadvis candidate
& \multirow{3}{0.4\textwidth}{Select events with $\ell$--$\tau$ pair candidate}
\\
Exactly one isolated light lepton &  \\
Opposite-sign charged $\ell$--$\tauhadvis$ pair &  \\
\midrule
$\taumT < \SI{35}{\GeV}$
& Reject \Ztt and \Wjets events \\
\midrule
$\mvis(\ell,\tauhadvis) > \SI{60}{\GeV}$ & Invariant mass of the $\ell$--$\tauhadvis$ pair. Reject events incompatible with $\ell$--$\tau$ pairs from $Z$-boson decays \\
\midrule
No tagged $b$-hadron jets
& Reject \ttbar and single-top-quark events \\
\midrule
Combined NN output $>$ 0.1 (0.2)
for events with 1P (3P) \tauhadvis candidates
& Reject background-like events \\
\midrule
NN (optimized for signal versus \Zll) output $>$ 0.2 & Ensure orthogonal region for correcting \Zll simulation
($\ell$ misidentified as 1P $\tauhadvis$ candidate, see Section~\ref{sec:modelling}) \\
\bottomrule
\end{tabular}
\end{table}
 
NN binary classifiers are used to distinguish signal events from \Wjets, \Ztt and \Zll background events.
The NNs are trained on simulated events (see Section~\ref{sec:modelling}).
Each individual NN is optimized to discriminate against a particular background process in a given decay channel.
The input to these NNs is a mixture of low-level and high-level
kinematic variables, as detailed in the Methods.
The low-level variables are the momentum components of the reconstructed $\ell$, \tauhadvis candidate and \MET.
The high-level variables are kinematic properties of the $\ell$--\tauhadvis--\MET system,
such as the collinear mass $\mcoll(\ell,\tau)$, defined as the invariant mass of the $\ell$--$\tauhadvis$--$\nu$
system, where the $\nu$ is assumed to have a momentum that is equal in \pT and $\phi$ to the measured $\MET$ and equal in $\eta$ to the
\tauhadvis momentum.
Given the finite training-sample size, the high-level variables help the NNs to converge faster,
while the NNs exploit any residual correlations between the low-level variables.
 
The outputs from the individual NNs are numbers between 0 and 1 that reflect the probability for an event to be a signal event;
they are combined into a final
discriminant, hereafter referred to as the `combined NN output'.
The combination is parameterized by weights associated with each individual NN and
optimized for discrimination among various background processes distributed differently along the range of combined NN output values, as detailed in the Methods.
This allows the maximum-likelihood fit to determine the background contributions more
precisely, which ultimately improves the sensitivity.
 
Events classified by the NNs as being background-like are excluded from the SR,
as indicated in Table~\ref{tab:region_defs}.
The signal acceptance times selection efficiency in the SR is
2.7\% for the $e\tau$ channel and
3.0\% for the $\mu\tau$ channel, as determined from simulated signal samples.

\section{Signal and background predictions}
\label{sec:modelling}
Predictions for signal and background contributions to the event yield
and kinematic distributions in the SR
are based partly on Monte Carlo (MC) simulations and partly on the use of
data in regions that are enriched in background events and do not overlap with the SR.
 
The signal events were simulated using \Pythia~8~\cite{Sjostrand:2014zea} with matrix elements calculated
at leading order (LO) in the strong coupling constant ($\alpha_\text{s}$).
Parameter values for initial-state radiation, multiparton interactions
and beam remnants were set according to the A14 set of tuned parameters (tune) \cite{ATL-PHYS-PUB-2014-021}
with the NNPDF~2.3 LO parton distribution function (PDF) set \cite{Ball:2012cx}.
Nominal signal samples were generated with
a parity-conserving $Z\ell\tau$ vertex and unpolarized $\tau$ leptons.
Scenarios where the decays
are maximally parity-violating were considered by reweighting the simulated events using \TauSpinner \cite{TauSpinner}. The event weight was computed as the probability of occurrence of each generated signal event, based on its kinematics, when assuming a specific $\tau$-polarization state (left-handed or right-handed).
 
Background \Ztt events were simulated with the \Sherpa~2.2.1~\cite{Bothmann:2019yzt} generator
using the NNPDF~3.0 NNLO PDF set~\cite{Ball:2014uwa}
and next-to-leading-order (NLO) matrix elements for up to two partons,
and LO matrix elements for up to four partons, calculated with the \Comix~\cite{Gleisberg:2008fv}
and \OpenLoops~\cite{Buccioni:2019sur,Cascioli:2011va,Denner:2016kdg} libraries.
They were matched with the \Sherpa parton shower~\cite{Schumann:2007mg} using the
MEPS@NLO prescription~\cite{Hoeche:2011fd,Hoeche:2012yf,Catani:2001cc,Hoeche:2009rj}
with the default \Sherpa tune.
This set-up follows the recommendations of the \Sherpa authors.
Background \Zll events were simulated using
the \PowhegBox~\cite{Alioli:2010xd} generator with NLO matrix elements
and interfaced to \Pythia~8 to model the parton showers, hadronization and underlying events.
All MC samples include a detailed simulation of the ATLAS detector
with $\GEANT4$ \cite{Agostinelli:2002hh},
to produce predictions that can be compared with the data.
Furthermore, simulated inelastic $pp$ collisions,
generated with \Pythia~8 using the NNPDF~2.3 LO PDF set and the A3 tune \cite{ATL-PHYS-PUB-2016-017},
were overlaid on the hard-scattering events to model the additional $pp$ collisions occurring in the same proton bunch crossing.
All simulated events were processed using the same reconstruction algorithms as used for data.
 
The simulation of $Z$-boson production is improved with a correction derived
from measurements in data. The simulated \pT spectra of the $Z$ boson are
reweighed to match the unfolded distribution measured by ATLAS in Ref.~\cite{STDM-2018-14}.
This improves the predictions of signal, \Ztt and \Zll events which are simulated
at different orders in $\alpha_\text{s}$ using different generators.
It also reduces the uncertainties related to missing higher orders in $\alpha_\text{s}$.
 
The predicted overall yields of signal and \Ztt events
are determined by a binned maximum-likelihood fit to data (see Section~\ref{sec:fit_and_result}) in the SR
and in a control region enhanced in $Z \to \tau\tau \to \ell\tauhadvis+3\nu$ events (\CRZtt),
using an unconstrained fit parameter,
which accounts for theoretical uncertainties in the total $Z$-boson production cross-section ($\sigma_{Z}$),
as well as the experimental uncertainties related to the acceptance of the common
$\ell\tauhadvis$ final state.
The selection criteria for
events in the \CRZtt are the same as those for events in the SR, except that events
are required to have $\taumT > \SI{35}{\GeV}$, $\lepmT < \SI{40}{\GeV}$, and
$\SI{70}{\GeV} < \mcoll(\ell, \tau) < \SI{110}{\GeV}$.
 
A much smaller contribution to the total background originates from \Zll events.
Their predicted overall yield is based on the measured value of $\sigma_{Z}$
\cite{STDM-2015-03} times the measured integrated luminosity.
The uncertainty in the measurement is taken into account.
The predicted rates of misidentifying electrons and muons in \Zll events as 1P \tauhadvis candidates
are corrected using data in a region enriched in \Zll events and orthogonal to the SR,
where the last selection criterion in Table~\ref{tab:region_defs} is inverted and the outputs of the NN classifiers optimized to reject \Ztt and \Wjets events are required to be greater than 0.8.
The corrections are derived as functions of \pT and $\abs{\eta}$ of the \tauhadvis candidate.
Statistical uncertainties in the correction are considered.

Events with fakes are one of the dominant contributions to the background, and are estimated
from data using the `fake-factor method', which is described in Ref.~\cite{EXOT-2016-36}.
A fake factor is defined as the ratio of the number of events with a fake \tauhadvis candidate passing the \Tight \tauhadvis identification requirement to those failing it.
Four fake factors, one for each of the most important backgrounds with fakes ($W(\to\ell\nu)\text{+jets}$, multijet, $Z(\to\ell\ell)\text{+jets}$ and \ttbar events), are measured in data in four corresponding fakes-enriched regions.
Each of these regions has a dominant contribution from one of the four targeted backgrounds with fakes.
These regions do not overlap with any of the regions used in the final maximum-likelihood fit.
The purity of the multijet-enriched region is improved by introducing two additional selection criteria: events must have a same-sign charged $\ell$--$\tauhadvis$ pair and $\lepmT < \SI{40}{\GeV}$.
The fake factors are measured as functions of the transverse momentum of the \tauhadvis candidate, separately for $e\tau$ and $\mu\tau$ events and for events with 1P or 3P \tauhadvis candidates.
 
The number of events with a fake 1P or 3P \tauhadvis candidate in a given \pT range in the SR or \CRZtt
is estimated by the number of events with a \tauhadvis candidate failing the \Tight identification requirement, but otherwise satisfying all other selection criteria for that region, multiplied by an average of the fake factors.
To calculate this average, the fake factors are summed with weights equal to the expected relative contribution of the corresponding background to the total yield of events in the region with the inverted identification requirement.
This approach is used to model the kinematic properties of the events with fakes.
The total predicted yields of these events in the SR and \CRZtt are instead
determined by a maximum-likelihood fit to data (see Section~\ref{sec:fit_and_result}), separately for events with 1P and 3P \tauhadvis candidates.
This approach avoids the uncertainties associated with the simulation of events with fakes, and makes full use of the large amount of data collected.
 
The remaining background processes (summarized as `Others' in the following) have relatively small contributions in
the SR and are estimated using simulations.
They include events from the production and decays of top quarks, pairs of gauge bosons, the Higgs boson and $W(\to\tau\nu)\text{+jets}$.
The yields of these events are normalized to their theoretical cross-sections.

\begin{figure}[p]
\centering
\includegraphics[width=0.9\textwidth]{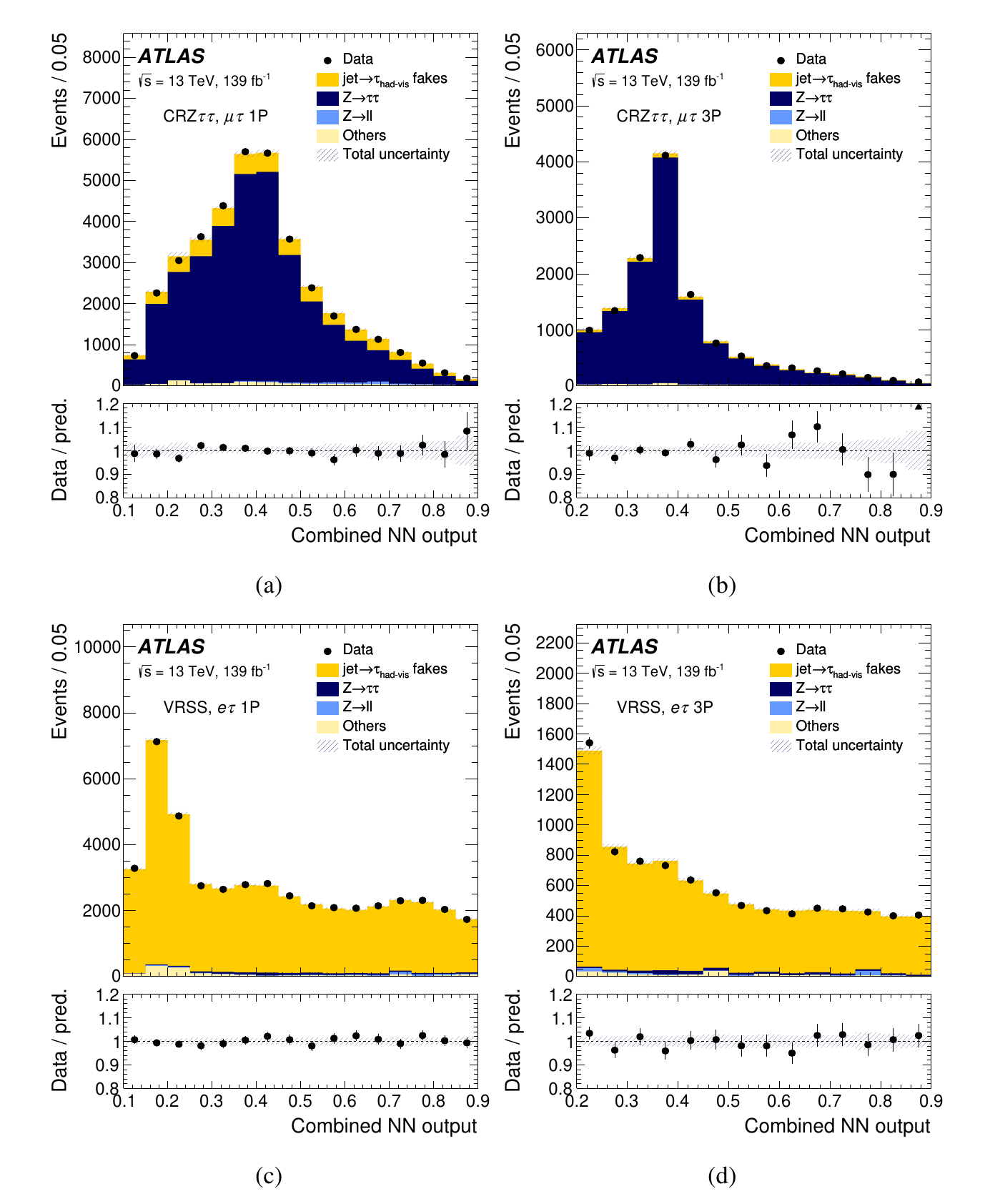}
\caption{
Distributions of the combined NN output in control regions and validation regions.
a,b, \CRZtt for the $\mu\tau$ channel with 1P (a) and 3P (b) \tauhadvis candidates.
c,d, Same-sign validation region (VRSS) for the $e\tau$ channel for events
with 1P (c) and 3P (d) \tauhadvis candidates.
The expected contributions are determined in a fit to data (Section~\ref{sec:fit_and_result}).
The panels below each plot show the ratios of the observed yields to the
best-fit background yields.
The hatched error bands represent a one standard deviation of the combined
statistical and systematic uncertainties.
The statistical uncertainties on the data are shown as vertical bars.
The last bin in each plot includes overflow events.
Similarly good agreement is observed in the same-sign validation region for
the $\mu\tau$ channel and \CRZtt for the $e\tau$ channel,
which are not shown here.}
\label{fig:postfit_dists_validations}
\end{figure}

The modelling of the estimated background is validated using events in regions where
a possible contamination from signal is negligible. Especially important to the search is the
modelling of the combined NN output distribution of \Ztt events and events with fakes. This
is validated by comparing the predicted distributions with data in the \CRZtt and
in a region similar to the SR, but with events that have same-sign charged $\ell$--$\tauhadvis$ pairs,  as shown in Figure~\ref{fig:postfit_dists_validations}.

\section{Constraints on $\BR(Z\to\ell\tau)$}
\label{sec:fit_and_result}
A statistical analysis of the selected events is performed to
assess the presence of LFV signal events.
The statistical analysis method is detailed in the Methods.
A simultaneous binned maximum-likelihood fit to the combined NN output in the SR and $\mcoll(\ell,\tau)$ in the \CRZtt is used to constrain uncertainties in the models and
extract evidence of a possible signal.
The fit is performed independently for the $e\tau$ and $\mu\tau$ channels.
Events with 1P and 3P \tauhadvis candidates are considered separately.
Hypothesis tests, in which a log-likelihood ratio is used as the test
statistic, are used to assess the compatibility between the background and signal models and the data.
 
There are four unconstrained parameters in the fits: two of them determine the
overall yields of events with fake 1P \tauhadvis or 3P
\tauhadvis candidates, one determines $\sigma_{Z}$ times the overall acceptance
and reconstruction efficiency of the $\ell\tauhadvis$ final state in $Z\to\tau\tau$ and signal events,
and the last one, the parameter of interest, determines the LFV branching fraction $\BR(Z\to\ell\tau)$ by modifying an arbitrary pre-fit signal yield.
 
Constrained parameters are also introduced to account for
systematic uncertainties in the signal and background predictions. In the case of no significant deviations from the SM background, exclusion limits are set using the \CLs method \cite{CLs}.
 
Systematic uncertainties in this search include uncertainties in simulated events in the modelling of trigger, reconstruction, identification and isolation
efficiencies, as well as energy calibrations and resolutions of reconstructed
objects.
Conservative theory uncertainties ranging between 4\% to 20\% are also assigned to
the predicted cross-sections used for the estimation of minor background processes.
These uncertainties are not assigned to events with fakes or $Z$-boson decays,
whose yields are determined from data.
These events constitute only a small fraction of the background events in the SR.
The dominant uncertainties in this search are those in the overall yields of
events with fakes, which are predominantly of statistical nature, and
those in the \tauhadvis energy calibration, which are
independent between 1P and 3P \tauhadvis candidates and
constrained by the fit
of the collinear mass spectrum to the data in the \CRZtt. A summary of the
uncertainties and their impact on the best-fit LFV branching fraction is given in Table~\ref{tab:syst_breakdown},
which shows that the sensitivity of the search is primarily limited by the available amount of data.
 
\begin{table}[!ht]
\centering
\caption{Summary of the uncertainties and their impacts on the measured signal branching fraction $\BR(\Zlt)$.
The statistical uncertainties include those in the determination of the yields of the events with fakes and from \Ztt or \Zlt decays.
The uncertainties related to light leptons include those in the trigger, reconstruction, identification and isolation
efficiencies, as well as energy calibrations.
The uncertainties related to jets and \MET include those in the energy calibration and resolution.
The uncertainties related to the $Z$-boson modelling include those in the correction of the simulated transverse momentum
and the measured production cross-section of the $Z$ boson.}
\label{tab:syst_breakdown}
\begin{tabularx}{0.95\textwidth}{lYY}
\toprule
& \multicolumn{2}{c}{\textbf{Uncertainty on $\BR(\Zlt)$ [$\times10^{-6}$]}} \\
\textbf{Source of uncertainty} & \textbf{$e\tau$} & \textbf{$\mu\tau$} \\
\midrule
Statistical                                         & $\pm$3.5 & $\pm$2.8 \\
Systematic                                          & $\pm$2.3 & $\pm$1.6 \\
\hspace{3ex} $\tau$ leptons                                    & $\pm$1.9 & $\pm$1.5 \\
\hspace{3ex}\hspace{3ex} Energy calibration         & $\pm$1.3 & $\pm$1.4 \\
\hspace{3ex}\hspace{3ex} Jet rejection              & $\pm$0.3 & $\pm$0.3 \\
\hspace{3ex}\hspace{3ex} Electron rejection         & $\pm$1.3 &          \\
\hspace{3ex} Light leptons                          & $\pm$0.4 & $\pm$0.1 \\
\hspace{3ex} \MET, jets and flavour tagging         & $\pm$0.6 & $\pm$0.5 \\
\hspace{3ex} $Z$-boson modelling               & $\pm$0.7 & $\pm$0.3 \\
\hspace{3ex} Luminosity and other minor backgrounds & $\pm$0.8 & $\pm$0.3 \\
\midrule
Total & $\pm$4.1 & $\pm$3.2 \\
\bottomrule
\end{tabularx}
\end{table}
 
\begin{figure}[p]
\centering
\includegraphics[width=0.9\textwidth]{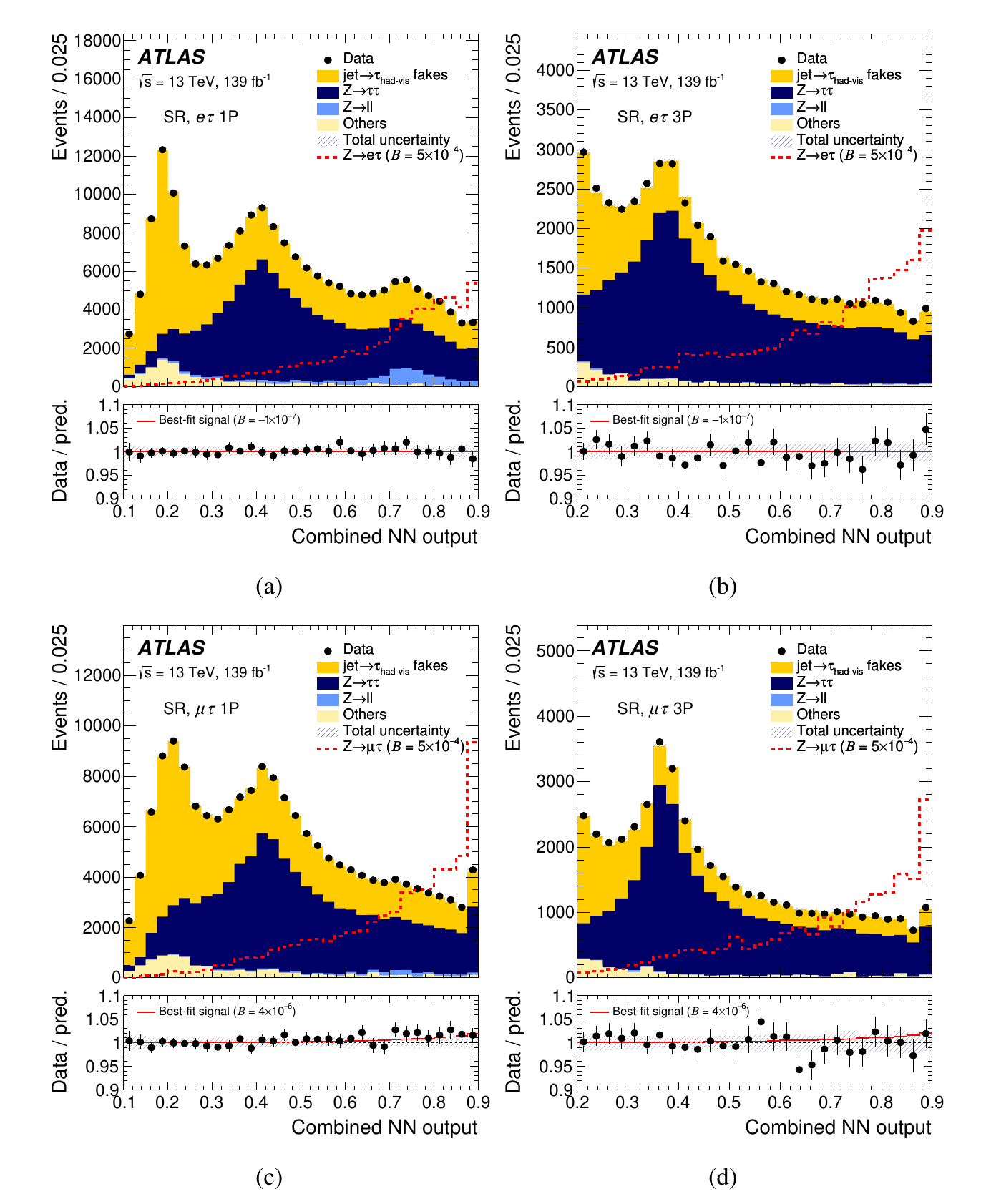}
\caption{
Distributions of the combined NN output in the signal region.
a,b, $e\tau$ events with 1P (a) and 3P (b) \tauhadvis candidates.
c,d, $\mu\tau$ events with 1P (c) and 3P (d) \tauhadvis candidates.
The expected contributions are determined in the fit to data.
The expected signal, normalized to $\BR(Z\to\ell\tau)=\num{5e-4}$, is shown as
a dashed red histogram in each plot.
The panels below each plot show the ratios of the observed yields (dots) and
the best-fit background-plus-signal yields (solid red line) to the best-fit
background yields.
The hatched error bands represent a one standard deviation of the combined
statistical and systematic uncertainties.
The statistical uncertainties on the data are shown as vertical bars.
The last bin in each plot includes overflow events.}
\label{fig:postfit_dists}
\end{figure}
 
\begin{table}[!ht]
\caption{Observed number of events and best-fit expected background and signal
yields in the SR. The additional requirement of combined NN output $>0.7$ is
included to consider the most signal-like events. The events with 1P and 3P \tauhadvis candidates are fitted simultaneously. The uncertainties include both the statistical and systematic contributions.}
\begin{center}
\begin{tabular}{l r@{\,\( \pm \)\,}l r@{\,\( \pm \)\,}l r@{\,\( \pm \)\,}l r@{\,\( \pm \)\,}l }
\toprule
& \multicolumn{2}{c}{\textbf{SR $e\tau$ 1P}} & \multicolumn{2}{c}{\textbf{SR $e\tau$ 3P}} & \multicolumn{2}{c}{\textbf{SR $\mu\tau$ 1P}} & \multicolumn{2}{c}{\textbf{SR $\mu\tau$ 3P}} \\
\midrule
Observed events & \multicolumn{2}{c}{35823} & \multicolumn{2}{c}{8108} & \multicolumn{2}{c}{27941} & \multicolumn{2}{c}{7462} \\
\midrule
Expected SM events & 35500 & 300          & 8120 & 90       & 27100 & 200          & 7600 & 90       \\
\midrule
Expected events with fakes & 13500 & 200          & 2400 & 90       & 9800 & 200          & 2010 & 70       \\
Expected $Z\to\tau\tau$ events  & 17100 & 200          & 5420 & 70   & 15600 & 200          & 5200 & 70           \\
Expected $Z\to\ell\ell$ events  & 4200 & 200          & 70 & 40       & 930 & 60          & 12.4 & 0.1          \\
Expected top-quark events & 130 & 10          & 30 & 4     & 120 & 10          & 44 & 6         \\
Expected $W(\to\tau\nu)$+jets events & 100 & 20          & 70 & 10        & 180 & 30          & 180 & 30      \\
Expected diboson events & 210 & 20          & 66 & 9       & 240 & 30          & 80 & 9        \\
Expected Higgs-boson events & 210 & 10          & 66 & 4       & 210 & 10          & 68 & 4          \\
\midrule
Pre-fit expected $Z\to \ell\tau$ events ($\BR=10^{-5}$) & 670 & 20 & 210 & 10 & 720 & 20 & 230 & 10\\
Best-fit $Z\to \ell\tau$ events & 0 & 300          & 0 & 80       & 300 & 200          & 90 & 70          \\
\bottomrule
\end{tabular}
\end{center}
\label{tab:postfit_highNN_SR_yields}
\end{table}
 
The best-fit expected and observed distributions of the combined NN output in
the SR are shown in Figure~\ref{fig:postfit_dists}.
The best-fit yields of \Ztt and events with fakes are close to the pre-fit predicted values
and are determined with a relative precision of 2\%--4\%.
Table~\ref{tab:postfit_highNN_SR_yields} shows
the best-fit expected background and signal yields and the observed number of events in the SR of the $e\tau$ and $\mu\tau$ channels
with an additional requirement of combined NN output $>0.7$ to consider the most signal-like events.
 
The best-fit amount of \Zlt signal corresponds to the branching fractions 
$\BR(\Zet)=(-0.1\pm 3.5\,(\text{stat}) \pm 2.3\,(\text{syst}))\times10^{-6}$ and $\BR(\Zmt)=(4.3\pm 2.8\,(\text{stat}) \pm 1.6\,(\text{syst}))\times10^{-6}$.
The positive best-fit value of $\BR(\Zmt)$ is related to a small excess of observed events relative to the background-only hypothesis.
This excess has a significance of 0.9 standard deviations when the events with 1P and 3P \tauhadvis candidates are fitted simultaneously.
 
No statistically significant deviation from the SM prediction is observed, and upper limits on the LFV branching fractions are set.
For the $\mu\tau$ channel, a more stringent upper limit is set by combining the likelihood function of the presented measurement and a similar measurement done with ATLAS Run 1 data \cite{HIGG-2015-09}.
Systematic uncertainties from the two measurements are considered uncorrelated in the combined likelihood function.
The upper limits are shown in Table~\ref{tab:ULs_unblind_tauSpin} for LFV decays with different assumptions about the $\tau$-polarization state.
In the scenario where the $\tau$ leptons are unpolarized, the observed upper limits at 95\% CL on $\BR(\Zet)$ and $\BR(\Zmt)$ are \num{8.1e-6} and \num{9.5e-6}, respectively.
 
\begin{table}[!ht]
\centering
\caption{Observed and expected (median) upper limits on the signal branching
fraction at 95\% CL. Upper limits for different $\tau$-polarization
scenarios are reported.
The differences between the observed and expected limits are due to the non-zero best-fit signal branching fractions.}
\label{tab:ULs_unblind_tauSpin}
\begin{tabularx}{\textwidth}{lYY}
\toprule
& \multicolumn{2}{c}{\textbf{Observed (expected) upper limit on $\BR(Z\to\ell\tau)$ [$\times10^{-6}$]}} \\
\textbf{Experiment, polarization assumption} & \textbf{$e\tau$} & \textbf{$\mu\tau$} \\
\midrule
ATLAS Run~2, unpolarized $\tau$  & 8.1 (8.1) & 9.9 (6.3) \\
ATLAS Run~2, left-handed $\tau$  & 8.2 (8.6) & 9.5 (6.7) \\
ATLAS Run~2, right-handed $\tau$ & 7.8 (7.6) & 10 (5.8) \\
\midrule
ATLAS Run~1, unpolarized $\tau$ \cite{HIGG-2015-09} &  & 17 (26) \\
\midrule
\makecell[l]{ATLAS Run~1+Run~2 combination,\\unpolarized $\tau$} &  & 9.5 (6.1) \\
\midrule\midrule
LEP OPAL, unpolarized $\tau$ \cite{LEP-Opal} & 9.8 & 17 \\
LEP DELPHI, unpolarized $\tau$ \cite{LEP-Delphi} & 22 & 12 \\
\bottomrule
\end{tabularx}
\end{table}
 
\FloatBarrier
In conclusion, these results from the ATLAS experiment at the LHC set stringent constraints on LFV $Z$-boson decays involving $\tau$ leptons (using only their hadronic decays), superseding the most stringent ones set by the LEP experiments more than two decades ago.
The precision of these results is mainly limited by statistical uncertainties.

\clearpage
\section*{Methods}
\subsection*{Neural network classifiers}
 
Several binary NN classifiers are trained for both the $e\tau$ and $\mu\tau$
channels to discriminate signal from the three major backgrounds: \Wjets, \Ztt
and \Zll. They are referred to as NN$_\text{Wjets}$, NN$_{\text{Z}\tau\tau}$ and NN$_{\text{Z}\ell\ell}$, respectively.
 
The NNs are trained using simulated events selected with the same criteria as those used in the SR,
except that the cuts on $\mvis(\ell,\tau)$ and the NN output are omitted,
and real \tauhadvis candidates from \Zlt and \Ztt are required to
pass less stringent identification criteria so as to increase the training sample size.
For the \Zll process, only events where the \tauhadvis
candidate is a misidentified light lepton are used.
For the \Wjets process, jets misidentified as \tauhadvis are modelled by simulations.
Different NNs are trained separately for $e\tau$ and $\mu\tau$ events as
well as for events with 1P or 3P \tauhadvis candidates.
To increase the signal sample size, the \Zet and \Zmt samples are combined and
used for training in both channels, assuming equivalent event topology when
exchanging $e$ and $\mu$.
Owing to the low expected yield of \Zll events with 3P \tauhadvis candidates,
no classifier is trained to discriminate them from background.

A mixture of low-level and high-level kinematic variables are used as input to the NNs.
The low-level variables include the four-momenta of the reconstructed
$\ell$ \cite{EGAM-2018-01,PERF-2015-10}, \tauhadvis candidate \cite{ATL-PHYS-PUB-2019-033,ATLAS-CONF-2017-029} and \MET \cite{PERF-2016-07,ATLAS-CONF-2018-023}.
To remove known spatial symmetries for optimal training, the low-level variables are transformed in
a way that preserves the Lorentz invariance before they are fed into the NNs.
The transformation consists of the following steps:
first, the $\ell$--\tauhadvis--\MET system is boosted in a direction in the plane
transverse to the beam line such that the total transverse momentum of the
system is zero;
the system is then rotated about the $z$ axis such that the direction of \MET is aligned
with the $x$ axis;
if the \tauhadvis candidate's momentum has a negative $z$ component, the entire system is
rotated about the new $x$ axis by \ang{180}.
After the transformation, only six independent non-vanishing components are left
(the \tauhadvis candidate is assumed to have zero rest mass), which are the inputs to the NNs.
 
The high-level variables include $\Delta\alpha$, which is a kinematic discriminant defined \cite{Davidson2012} as
\begin{equation}\nonumber
\Delta\alpha = \frac{m_Z^2 - m_\tau^2}{2 p(\ell)\cdot p(\tauhadvis)}
- \frac{\pT(\ell)}{\pT(\tauhadvis)},
\end{equation}
where $m_Z$ and $m_\tau$ are the nominal masses of the $Z$ boson and $\tau$ lepton,
respectively, and $p$ denotes four-momentum.
It is specifically defined to test the assumptions that the missing momentum of
the event is collinear with the \tauhadvis candidate, and that the $\tau$ and
light leptons in the event are decay products of an on-shell $Z$ boson.
For a signal event, where these assumptions are approximately true, it is
expected that $\Delta\alpha\approx0$. Meanwhile, for a SM background event, the
value is expected to deviate from zero in general.
The other high-level variables are the invariant mass of the $\ell-\tauhadvis$ system,
the collinear mass $\mcoll(\ell,\tau)$ and the invariant mass of the light lepton and
the track associated with the $\tauhadvis$ candidate
(only used by the $Z\to\ell\ell$ classifier).
 
The training and optimization of the NN classifiers are performed using the
open-source software package \Keras \cite{chollet2015keras}. All of the NNs used in the
analysis share the same architecture. Each NN consists of an input layer, two
hidden layers of 20 nodes each, and an output layer with a single node.
Each layer is fully connected to the neighbouring layers. Low-level and
high-level variables are treated in the same way in the input layer.  The hidden-layer nodes use rectified linear activation functions, while the output node uses a sigmoid activation function.  The NNs are trained using the Adam algorithm
\cite{Adam} to optimize the binary cross entropy. All the NNs are trained with a
batch size of 256 and 200 epochs. The number of hidden layers, the number of
nodes per layer, the training batch size and the learning rate parameter of the
optimizer are simultaneously chosen by maximizing the area under the expected
receiver operating characteristic curve. The optimization is done
with a grid scan. No regularization or dropout is added, and no sign of
overtraining is observed. For other configurations and hyperparameters
that have not been mentioned, the default settings in \Keras 1.1.0 are used.
 
Each NN classifier outputs a score between 0 and 1 for each event, where a
higher score indicates that the event is more signal-like. The output
scores from the different classifiers are combined into the final discriminant
(combined NN output) using the formula
\begin{equation}\nonumber
\text{combined NN output} = 1 - \sqrt{\frac{\sum_b w_b \times
\left(1 - \text{NN}_b\text{ output}\right)^2}{\sum_b w_b}},
\end{equation}
where $b=\text{Wjets},\text{Z}\tau\tau,\text{Z}\ell\ell$
and $w_b$ are constant parameters.
Output scores for events with 1P \tauhadvis
candidates and those with 3P \tauhadvis candidates are combined separately.
The summation is over Wjets, Z$\tau\tau$ and Z$\ell\ell$ for events with 1P \tauhadvis
candidates, and only over Wjets and Z$\tau\tau$ for events with 3P \tauhadvis
candidates.
 
By construction, the combined NN output ranges between 0 and 1, where
0 represents the most background-like (and 1 the most signal-like) event
possible.
The choice of values of $w_b$ affects the expected sensitivity of
the analysis because they change how events from the different background processes are distributed
along the range of combined NN output values, and thus impacts the ability of the binned
maximum-likelihood fit to determine the  background contributions. The values of $w_b$ are chosen with a grid scan to
minimize the expected upper limit on the branching fraction in the absence of a signal. The chosen
values have the ratio
$w_{\text{Z}\tau\tau} : w_{\text{Wjets}} : w_{\text{Z}\ell\ell} = 1.0 : 1.5 : 0.33$.
As could be expected, the optimized weights loosely reflect the impact of the
uncertainties in the corresponding backgrounds on the determination of the signal branching fraction.

\subsection*{Maximum-likelihood fit}
 
Binned maximum-likelihood fits are implemented using the statistical
analysis packages \textsc{RooFit} \cite{RooFit}, \textsc{RooStats} \cite{RooStats}
and \textsc{HistFitter}~\cite{HistFitter}.
The expected binned distributions of the combined NN output in the SR and the collinear
mass in the \CRZtt are fit to data to extract evidence of signal events.
Fitting the data in the \CRZtt and in part of the SR with low combined NN output values (where no signal is expected)
benefits the overall sensitivity to the signal, because it reduces the uncertainties of the
background model in the high combined NN output value region, where
most of the signal is expected.
Owing to the differences in background composition, acceptance and efficiencies,
regions with 1P and 3P \tauhadvis candidates are fit separately but
simultaneously. The probabilities of compatibility between the data and the
background-only or background-plus-signal hypotheses are assessed using the modified frequentist
\CLs method~\cite{CLs}, and exclusion upper limits on
$\BR (Z\to\ell\tau)$ are set by the inversion of these hypothesis tests.
 
The background-plus-signal model has four unconstrained parameters before the fit.
Two of the parameters determine the overall yields of events with 1P and 3P fakes separately.
A third parameter determines $\sigma_{Z}$ times the
overall acceptance and reconstruction efficiency of events with a true
$\ell\tauhadvis$ final state.
It is applied to the normalizations of both the signal and \Ztt events to ensure
that the same $\sigma_{Z}$ times acceptance is estimated for both processes.
The last unconstrained parameter is the parameter of interest \musig, which controls the normalization of signal events.
Given the similarity between the signal and $Z \to \tau\tau \to \ell\tauhadvis + 3\nu$ final states and that both processes are estimated with the same $\sigma_{Z}$ and acceptance and efficiency corrections,
this choice of parameterization reduces the impact on the determined $\BR(Z\to\ell\tau)$ from detector effects and uncertainties in predicting
$\sigma_{Z}$.
The parameter of interest represents
\begin{equation}\nonumber
\musig = \frac{\BR(Z\to\ell\tau)}{\BR_\text{pre-fit}(Z\to\ell\tau)},
\end{equation}
where $\BR_\text{pre-fit}(Z\to\ell\tau)$ is an arbitrary branching fraction
to which the signal prediction is normalized.  Although the physical branching fraction must be positive,  the parameter of interest in the fit is not constrained to be positive.
 
Systematic uncertainties are represented by nuisance parameters (NPs) with Gaussian
constraints in the likelihood function.
The impact of uncertainties on both the shape and normalization of the predicted distributions are taken into account.
Uncertainties in the energy calibration and resolution, and in the trigger,
reconstruction, identification and isolation efficiencies of jets, electrons,
muons, \tauhadvis and \MET are considered.
Theoretical uncertainties in the production cross sections affect only the
predictions of the minor backgrounds, because the \Ztt and signal yields are determined in the
maximum-likelihood fit to data and the \Zll yield is determined by the measured
value of $\sigma_{Z}$.
Statistical uncertainties in the determination of the fake factors are also
considered. They are modelled by one NP per \pT bin in which the fake factors are measured.
As noted in Section \ref{sec:fit_and_result}, the dominant uncertainties in the analysis are
the statistical uncertainties in determining how many events have fakes
and the systematic uncertainties in the reconstructed $\tauhadvis$ energy.

For the $\mu\tau$ channel, the likelihood functions of the presented measurement
and of the measurement in Ref.~\cite{HIGG-2015-09} are combined.
As the two measurements are statistically uncorrelated and the predictions are
based on different methods, NPs in the individual likelihood
functions are considered uncorrelated in the combination. The method of
combination is the same as in Ref.~\cite{EXOT-2016-36}.

\FloatBarrier
\clearpage
 
\section*{Acknowledgements}

We wish to acknowledge our late colleague, Olga Igonkina (1973--2019), for inspiring and driving this and other searches for lepton flavour violation within the ATLAS experiment.
Her curiosity and intelligence remain an inspiration to the ATLAS Collaboration.
 

We thank CERN for the very successful operation of the LHC, as well as the
support staff from our institutions without whom ATLAS could not be
operated efficiently.
 
We acknowledge the support of ANPCyT, Argentina; YerPhI, Armenia; ARC, Australia; BMWFW and FWF, Austria; ANAS, Azerbaijan; SSTC, Belarus; CNPq and FAPESP, Brazil; NSERC, NRC and CFI, Canada; CERN; ANID, Chile; CAS, MOST and NSFC, China; COLCIENCIAS, Colombia; MSMT CR, MPO CR and VSC CR, Czech Republic; DNRF and DNSRC, Denmark; IN2P3-CNRS and CEA-DRF/IRFU, France; SRNSFG, Georgia; BMBF, HGF and MPG, Germany; GSRT, Greece; RGC and Hong Kong SAR, China; ISF and Benoziyo Center, Israel; INFN, Italy; MEXT and JSPS, Japan; CNRST, Morocco; NWO, Netherlands; RCN, Norway; MNiSW and NCN, Poland; FCT, Portugal; MNE/IFA, Romania; MES of Russia and NRC KI, Russia Federation; JINR; MESTD, Serbia; MSSR, Slovakia; ARRS and MIZ\v{S}, Slovenia; DST/NRF, South Africa; MICINN, Spain; SRC and Wallenberg Foundation, Sweden; SERI, SNSF and Cantons of Bern and Geneva, Switzerland; MOST, Taiwan; TAEK, Turkey; STFC, United Kingdom; DOE and NSF, United States of America. In addition, individual groups and members have received support from BCKDF, CANARIE, Compute Canada and CRC, Canada; ERC, ERDF, Horizon 2020, Marie Sk{\l}odowska-Curie Actions and COST, European Union; Investissements d'Avenir Labex, Investissements d'Avenir Idex and ANR, France; DFG and AvH Foundation, Germany; Herakleitos, Thales and Aristeia programmes co-financed by EU-ESF and the Greek NSRF, Greece; BSF-NSF and GIF, Israel; La Caixa Banking Foundation, CERCA Programme Generalitat de Catalunya and PROMETEO and GenT Programmes Generalitat Valenciana, Spain; G\"{o}ran Gustafssons Stiftelse, Sweden; The Royal Society and Leverhulme Trust, United Kingdom.
 
The crucial computing support from all WLCG partners is acknowledged gratefully, in particular from CERN, the ATLAS Tier-1 facilities at TRIUMF (Canada), NDGF (Denmark, Norway, Sweden), CC-IN2P3 (France), KIT/GridKA (Germany), INFN-CNAF (Italy), NL-T1 (Netherlands), PIC (Spain), ASGC (Taiwan), RAL (UK) and BNL (USA), the Tier-2 facilities worldwide and large non-WLCG resource providers. Major contributors of computing resources are listed in Ref.~\cite{ATL-SOFT-PUB-2020-001}.
 
 
\clearpage
\section*{Extended Data}
 
\newcounter{originalfigure}
\setcounter{originalfigure}{\value{figure}}
\setcounter{figure}{0}
 
\begin{figure}[!h]
\centering
\includegraphics[width=\textwidth]{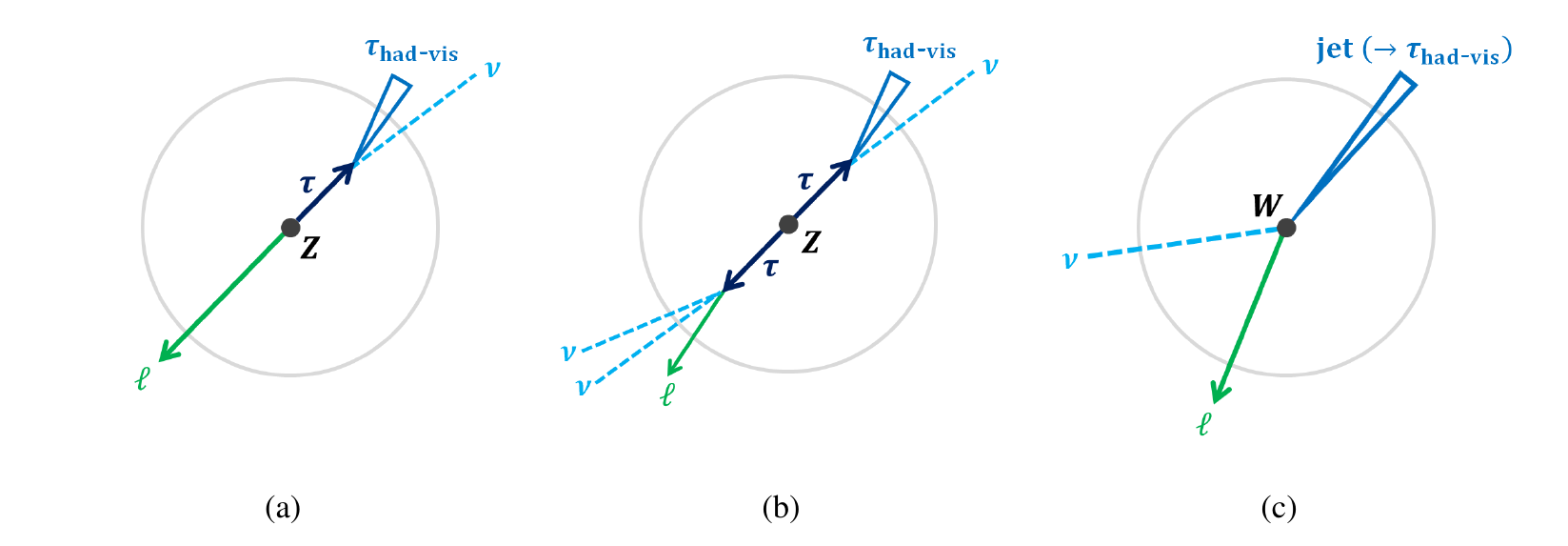}
\caption{
Schematic representation of a typical event selected in the SR.
The topology as seen in the plane transverse to the beam line is shown.
(a)~A signal \Zlt event.
(b)~A \Ztt event.
(c)~A \Wjets event.
The green arrows represent reconstructed light leptons ($\ell$).
The blue triangles represent the \tauhadvis candidates.
The light blue dashed lines represent neutrinos that
escape detection and are reconstructed as (part of) the missing transverse
momentum of the event. }
\label{fig:mT_dists}
\end{figure}
 
\begin{figure}[!h]
\centering
\includegraphics[width=\textwidth]{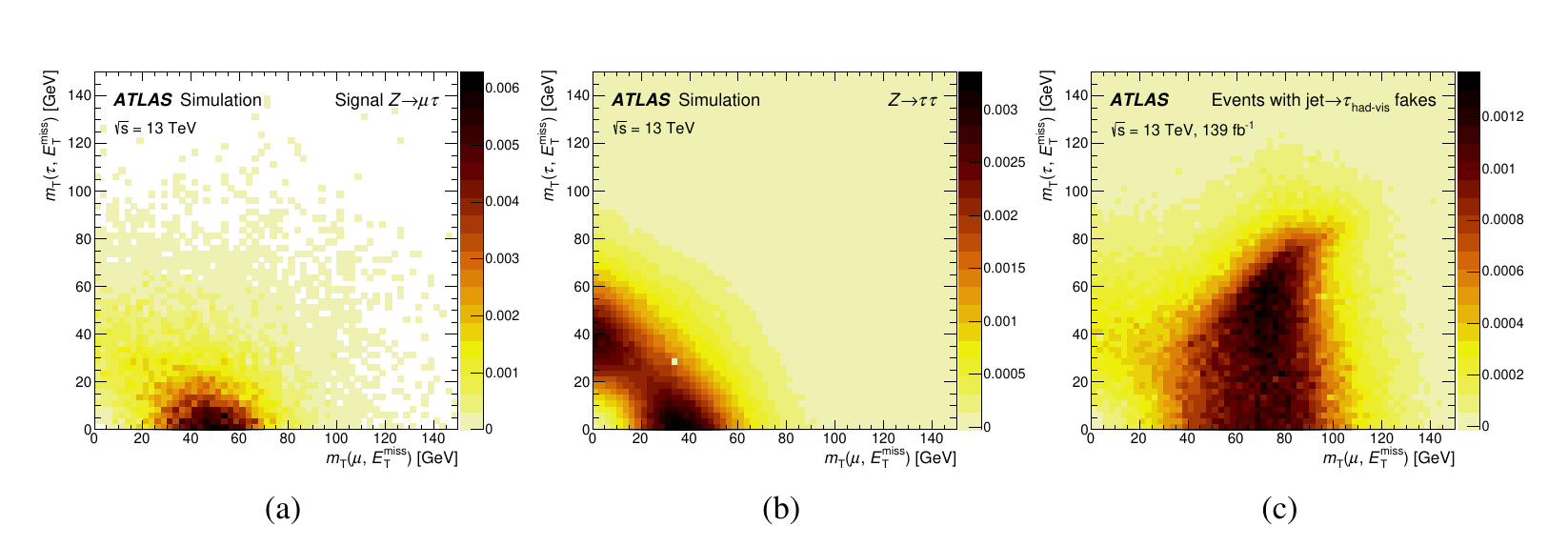}
\caption{
Distributions of \taumT versus \mumT of events selected in the SR.
(a)~Simulated \Zmt events.
(b)~Simulated \Ztt events.
(c)~Events measured in data in regions where
quark- or gluon-initiated jets are misidentified as \tauhadvis candidates
(events with jet$\to\tauhadvis$ fakes, see Section~\ref{sec:modelling})
in the $\mu\tau$ final state.
The colour map represents the fraction of events in each bin.}
\label{fig:mT_dists_2D}
\end{figure}
 
\setcounter{figure}{\value{originalfigure}}
 
\clearpage

\printbibliography
 
\clearpage
 
\begin{flushleft}
\hypersetup{urlcolor=black}
{\Large The ATLAS Collaboration}

\bigskip

\AtlasOrcid[0000-0002-6665-4934]{G.~Aad}$^\textrm{\scriptsize 102}$,    
\AtlasOrcid[0000-0002-5888-2734]{B.~Abbott}$^\textrm{\scriptsize 128}$,    
\AtlasOrcid{D.C.~Abbott}$^\textrm{\scriptsize 103}$,    
\AtlasOrcid[0000-0002-2788-3822]{A.~Abed~Abud}$^\textrm{\scriptsize 36}$,    
\AtlasOrcid[0000-0002-1002-1652]{K.~Abeling}$^\textrm{\scriptsize 53}$,    
\AtlasOrcid[0000-0002-2987-4006]{D.K.~Abhayasinghe}$^\textrm{\scriptsize 94}$,    
\AtlasOrcid[0000-0002-8496-9294]{S.H.~Abidi}$^\textrm{\scriptsize 167}$,    
\AtlasOrcid[0000-0002-8279-9324]{O.S.~AbouZeid}$^\textrm{\scriptsize 40}$,    
\AtlasOrcid{N.L.~Abraham}$^\textrm{\scriptsize 156}$,    
\AtlasOrcid[0000-0001-5329-6640]{H.~Abramowicz}$^\textrm{\scriptsize 161}$
% ...
\bigskip
\\

% ...
$^{36}$CERN, Geneva; Switzerland.\\
% ...
$^{40}$Niels Bohr Institute, University of Copenhagen, Copenhagen; Denmark.\\
% ...
$^{53}$II. Physikalisches Institut, Georg-August-Universit\"{a}t G\"ottingen, G\"ottingen; Germany.\\
% ...
$^{94}$Department of Physics, Royal Holloway University of London, Egham; United Kingdom.\\
% ...
$^{102}$CPPM, Aix-Marseille Universit\'e, CNRS/IN2P3, Marseille; France.\\
$^{103}$Department of Physics, University of Massachusetts, Amherst MA; United States of America.\\
% ...
$^{128}$Homer L. Dodge Department of Physics and Astronomy, University of Oklahoma, Norman OK; United States of America.\\
% ...
$^{156}$Department of Physics and Astronomy, University of Sussex, Brighton; United Kingdom.\\
% ...
$^{161}$Raymond and Beverly Sackler School of Physics and Astronomy, Tel Aviv University, Tel Aviv; Israel.\\
% ...
$^{167}$Department of Physics, University of Toronto, Toronto ON; Canada.\\
% ...

\end{flushleft}


\end{document}